\documentclass{article}

\PassOptionsToPackage{numbers, compress}{natbib}
\usepackage[preprint]{neurips_2025}

\usepackage[utf8]{inputenc}  
\usepackage[T1]{fontenc}     

\usepackage{xcolor}
\definecolor{darkpastelblue}{rgb}{0.66, 0.13, 0.24}
\usepackage[colorlinks, anchorcolor=darkpastelblue,
    linkcolor=darkpastelblue, urlcolor=darkpastelblue,
citecolor=darkpastelblue]{hyperref}

\usepackage{url}        
\usepackage{booktabs}   
\usepackage{amsfonts}   
\usepackage{nicefrac}   
\usepackage{microtype}  

\usepackage[american]{babel}   
\usepackage{graphicx}          
\usepackage{subcaption}        
\usepackage{amsmath, amssymb}  
\usepackage{physics}           
\usepackage{cleveref}          

\title{Toward the Thermodynamic Limit: Neural Operators for Non-equilibrium Dynamics of Mott Insulators}

\author{
    Miles Waugh$^*$ \\
    University of California, Irvine \\
    \texttt{waughmm@uci.edu} \\
    \And
    Chuwei Wang$^*$ \\
    California Institute of Technology \\
    \texttt{chuweiw@caltech.edu} \\
    \And
    Radu Andrei \\
    ETH Z\"urich \\
    \texttt{randrei@student.ethz.ch} \\
    \And
    Nusair Islam \\
    NVIDIA Research \\
    \texttt{nislam@nvidia.com} \\
    \And
        Filip Marijanović \\
    ETH Z\"urich \\
    \texttt{fmarijanovic@student.ethz.ch} \\
    \And
    Taylor Patti \\
    NVIDIA Research \\
    \texttt{tpatti@nvidia.com} \\
    \And
    Eugene Demler \\
    ETH Z\"urich \\
    \texttt{edemler@ethz.ch} \\
    \And
    Anima Anandkumar \\
    California Institute of Technology \\
    \texttt{anima@caltech.edu} \\
}

\begin{document}

\maketitle
\footnotetext[1]{\hspace{-0.5em}$^{*}$Equal contribution.}

\begin{abstract}
    Mott insulators exhibit complex photoexcitation dynamics under intense optical driving, with potential implications for carrier multiplication beyond the Shockley–Queisser limit. Probing these nonequilibrium processes requires access to the thermodynamic limit, where the number of lattice sites becomes arbitrarily large, but conventional solvers are constrained to small systems due to the exponential growth of the Hilbert space. Fourier Neural Operators (FNOs), originally developed for solving partial differential equations, naturally accommodate inputs of varying resolution and are capable of capturing nonlocal effects. Here, we employ FNOs to learn the mapping from noise-perturbed ground-state momentum distributions to their post-pulse counterparts across a range of interaction strengths and driving parameters. Trained only on small lattices, the model generalizes zero-shot to much larger systems, producing physically reasonable momentum distributions well beyond the reach of numerical solvers.
    Specifically, the model can predict momentum distribution for a $1024^2$ system within a few seconds that matches the theoretical behavior of key observables, whereas direct numerical simulations have so far been restricted to edge sizes of $\sim 30$.
    These results demonstrate the potential of neural operators to directly access large-scale nonequilibrium dynamics, providing a new pathway toward the thermodynamic limit in strongly correlated materials.
\end{abstract}

\vspace{-0.8em}
\section{Introduction}
\vspace{-0.8em}

Quantum many-body simulations of strongly correlated systems represent one of the most computationally demanding challenges in condensed matter physics, with applications spanning quantum magnetism~\cite{anderson_antiferromagnetism_1950, balents_spin_2010} and high-temperature superconductivity~\cite{lee_doping_2006, sakai_direct_2018}.
The two-dimensional single-band Hubbard model stands as a minimal yet rich platform for studying Mott insulating behavior, antiferromagnetism, and photoinduced carrier dynamics relevant to emergent optoelectronic phenomena~\cite{imada1998metal,kaneko2019photoinduced,mehio2023hubbard}.
Because finite-size effects can alter spectral and transport behaviors, and some collective behaviors only emerge at macroscopic scales, accessing these phenomena quantitatively in experimentally relevant regimes requires simulations on lattice sizes that approach the thermodynamic limit $N\to\infty$, where $N$ is the linear system size, such that there are $N^2$ total lattice sites.

However, this limit has remained largely inaccessible to exact numerical methods due to the exponential growth of the Hilbert space dimension w.r.t. $N^2$, which makes even modest system sizes computationally intractable~\cite{sandvik2010computational,schollwock2011density}. The challenge is amplified in strongly correlated regimes, where perturbative approaches fail and the full many-body problem must be tackled directly~\cite{georges1996dynamical}. External driving fields, essential for modeling realistic photoexcitation experiments, introduce additional time-dependent complexity since the simulation requires fine temporal discretization to capture rapid field-induced variations~\cite{suzuki1990fractal}. 

Recent advances have demonstrated the promise of machine learning approaches in quantum-related problems~\cite{ carleo_solving_2017,gu_solving_2025, zhang_complex-valued_2025, shah_fourier_2024, zhang_neural_2025, mizera_scattering_2023, li2024computational}. However, most studies address specific systems individually, with limited exploration of transferability and scale-up across different system sizes. Even when generalization is considered, it typically involves interpolation within a restricted range rather than the extrapolation required here~\cite{foster2025ab,gerard2024transferable, scherbela2023variational}. Neural operators~\cite{kovachki2023neural} and transformers~\cite{vaswani2017attention} represent two state-of-the-art architectures that accept size-varying inputs and capture nonlocal interactions, both of which are highly favorable properties for strongly correlated quantum systems with nonlocal entanglement. They have widespread applications in physical sciences, especially for solving PDEs~\cite{mizera_scattering_2023, zhang_complex-valued_2025, wang2024beyond,wu2024transolver,tolooshams2025equireg,li2022transformer,wang2025accelerating,bhojwani2025black}. Yet transferability across system sizes in quantum lattices remains underexplored. In practice, heavy numerical methods are limited to very small lattices, where strong finite-size artifacts, decaying slowly as $1/N$, can obscure the underlying physics. Furthermore, in the case of scattering processes, momentum and energy conservation impose strong constraints on possible final states, therefore requiring sufficiently large system sizes to reasonably approach the continuum description.
Related works are discussed in \Cref{apdxA:rel_work}.



This work addresses a physically motivated question: can models trained exclusively on small lattices display zero-shot transferability to much larger systems, thereby providing a practical route toward the thermodynamic limit? In this work, we focus on momentum distributions, which play a vital role in many downstream tasks, including transport properties, optical conductivity, and photoemission spectroscopy~\cite{valli2012correlation,lee2006doping,imada1998metal,datta1997electronic}.

We test FNO~\cite{li2020fourier} and ViT~\cite{dosovitskiy2020image} architectures, both tailored for this quantum lattice problem, in mapping initial ground states with different noise realizations to post-pulse momentum distributions across varying pulse parameters, interaction ratios $U/\tau$, and system sizes $N$. We train the models on systems up to $N = 20$
(the largest size for which we can conduct numerical simulations to produce a sufficient amount of data for training and validation).
While both architectures show similar accuracy within the training regime, FNO produces physically reasonable results for $N$ as large as 2048, where ground truth is not available, whereas ViT exhibits unphysical artifacts, establishing FNO as a potential approach to target the thermodynamic limit.

\vspace{-0.6em}
\section{Method}
\vspace{-0.6em}
\subsection{Problem Formulation}

We study the nonequilibrium dynamics of strongly correlated electrons on a two-dimensional square lattice, described by the single-band Hubbard model~\cite{hubbard1963electron}. The time-dependent Hamiltonian under external driving is
$    H(t) = -\tau\sum_{\mathbf{j},\boldsymbol{\delta},\alpha} e^{ie\boldsymbol{\delta\cdot A}(t)} c_{\mathbf{j}+\boldsymbol{\delta},\alpha}^{\dagger} c_{\mathbf{j},\alpha} + U \sum_{\mathbf{j}} n_{\mathbf{j}\uparrow} n_{\mathbf{j}\downarrow},$
where $c_{\mathbf{j},\alpha}^{(\dagger)}$ are fermionic annihilation (creation) operators at site $\mathbf{j}\in {\{1,2,\ldots,N\}}^2$ with spin $\alpha \in \{\uparrow,\downarrow\}$, $\boldsymbol{\delta} \in \{\pm\hat{x}, \pm\hat{y}\}$ connects nearest neighbors, $\tau$ is the hopping amplitude, and $U$ represents the on-site Coulomb repulsion. The $e$ in the exponent denotes electron charge. External optical pulses are incorporated via Peierls substitution through the time-dependent vector potential $\mathbf{A}(t)$. We employ Gaussian pulses $A(t)=A_0\exp\left(-4\ln(2)\frac{{(t-t_0)}^{2}}{\tau_{\text{FWHM}}^2}\right)\cos(\omega(t-t_0))$,
where $A_0$, $\omega$, and $\tau_{\text{FWHM}}$ control the amplitude, frequency, and width of the drive.

The evolution of a quantum system can be described by the Schrödinger equation $i\pdv{t}\ket{\psi(t)} = H(t) \ket{\psi(t)}$. In our setting, the state is characterized by the correlation matrix.
To mimic fluctuations and imperfections present in realistic systems, the simulations start from the ground-state correlation matrix perturbed by random noise.
The computational complexity of correlation-matrix propagation scales as $O(N^6 T)$, where $T$ is the number of time steps.
In this work, we focus on predicting the post-pulse momentum distribution $ n(\mathbf{k,}t)$ with $\mathbf{k}$ spanning the first Brillouin zone. This is a 2D observable of great physical importance and with vast applications in downstream tasks. Please refer to \Cref{apdx:phy} for more physical background, the specific setting in this study, and the motivation for predicting non-equilibrium post-pulse momentum distribution.

\begin{figure}[t]
    \centering

    \includegraphics[width=1\linewidth]{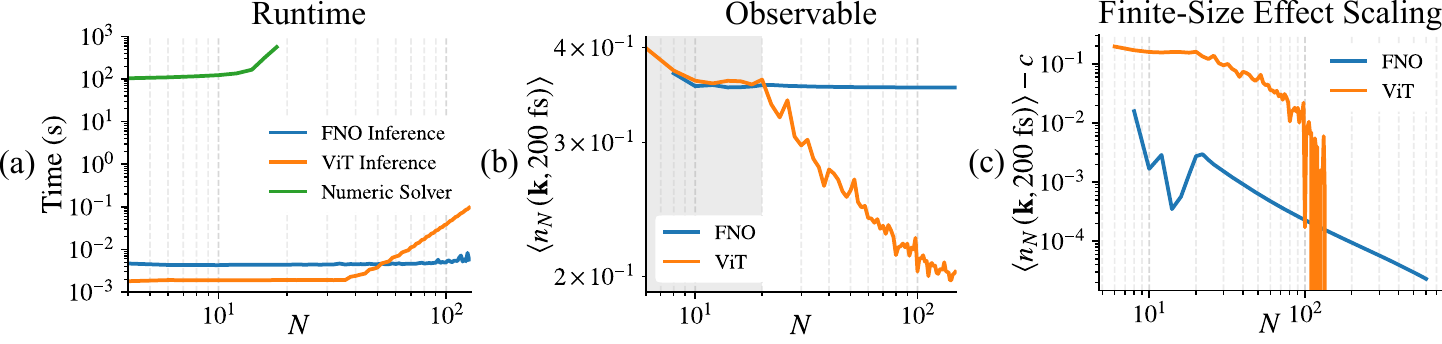}

    \caption{\@
        \textbf{(a)} Runtime of the numerical solver, FNO inference, and ViT inference for one noise realization on an Nvidia RTX 5080 GPU.\@
        \textbf{(b)} The predicted average momentum density, $\langle n_{N}\left(\textbf{k}, 200\text{ fs}\right)\rangle$, as a function of the system size, $N$. The FNO correctly stabilizes toward a limiting value as $N\to\infty$, while the ViT does not. Here, $U/\tau=3.497$, and a pulse of $A=0.5$ MV/cm and $\omega=10$ THz is used. An approximate initial state is used due to the computational limitations of the numerical simulation.
        \textbf{(c)} The difference between the predicted average momentum density from (b) and its approximate limiting value, $c$, as $N$ becomes large. For the ViT, this limiting value is the average momentum density at $N=148$. For the FNO, this limiting value is the average momentum density at $N=2048$. The FNO converges stably toward a limiting value with a power law scaling.
    }\label{fig:figure_1}
    \vspace{-0.8em}
\end{figure}

\vspace{-0.5em}
\subsection{Efficient Data-driven Transferable Models}
\vspace{-0.5em}
In this work, we explore data-driven methods to efficiently predict the post-pulse momentum distribution. Specifically, the networks take as input the initial momentum distribution at $t=0$, as well as the system configurations, the $U/\tau$ ratio, the reciprocal system size $1/N$ (where the system is $N\times N$), the pulse amplitude $A_0$, and the pulse frequency $\omega$.
We require the model to be able to deal with inputs of different system sizes, as we will ultimately study the limit as $N\to\infty$. Therefore, we choose two state-of-the-art architectures with this property, Fourier Neural Operator (FNO)~\cite{li2020fourier} and (vision-) transformer (ViT)~\cite{vaswani2017attention,dosovitskiy2020image}, as the backbone.

With ground-truth data obtained from numerical simulation on various small system sizes (up to $N=20$) and pulse parameters, we train the models to learn the following mapping,
\begin{equation}
    \mathcal{G}_\theta: {\left\{\left(n\left(\textbf{k},0\right),\eta,\texttt{PE}\left(\textbf{k}\right)\right)\right\}}_{\textbf{k}} \mapsto {\left\{n\left(\textbf{k},T\right)\right\}}_{\textbf{k}},
\end{equation}
where $\eta = \left(U/\tau,A_0,\omega,1/N\right)$ are the system and pulse parameters, $\textbf{k}=\left(k_{x},k_{y}\right)\in{\left(-\pi,\pi\right]}^2$ are the allowed momenta in the Brillouin zone, i.e. $(\frac i N\pi, \frac j N \pi)$ for $(i,j)\in\mathbb{Z}^2$, $\texttt{PE}$ denotes the position embedding of the momentum grid $\textbf{k}$, and $T=200$ fs. Implementation details are shown in appx. \ref{apdx_imple}.

\vspace{-0.5em}
\paragraph{Fourier Neural Operator}
\vspace{-0.5em}
Unlike traditional neural networks, neural operators, originally developed to solve PDEs, learn mappings between function spaces. Fourier Neural Operators have proven effective for learning solution operators for PDEs and can efficiently learn global interactions. In lattice systems, it is not always evident or justified that a finite-$N$ lattice can be regarded merely as a discretization of a continuum system, motivating the need to explicitly provide the system size as an input to the model. In the previous example, we provided the model with the information of the system size through a channel with a constant $1/N$. We remark that, as shown in our ablation study (\Cref{apdx_abl}), the model is able to capture the relation with $N$ even without explicit physical guidance, and can adapt to generic forms such as $N^{-\alpha}$ which approach 0 as $N\to\infty$.

\vspace{-0.5em}
\paragraph{Vision Transformer}
\vspace{-0.5em}
Vision transformers have demonstrated strong capability in extracting both global features and fine details of images. Input images are divided into non-overlapping patches of fixed size, each treated as a token analogous to words in a language sequence, so that the sequence length scales with the system size. The use of patches, however, imposes a constraint that the lattice size $N$ must be divisible by the patch size. To ensure broad applicability across arbitrary $N$, we primarily consider patch sizes of 1 and 2, while the effects of larger patches are discussed in our ablation study (\Cref{apdx_abl}).

\section{Experimental Results}
The models are trained on 900 samples and validated on 100 samples for each system size in $\left\{6,8,\ldots,20\right\}$. The center and width of the second resonant pulse are chosen as $t_0=100$ fs and $\tau_{\mathrm{FWHM}}=\frac{100}{\sqrt{\ln 4}}$ fs, such that 95\% of the pulse is contained across the timespan $t=[0,200]$ fs. We trained the model on $U/\tau\in[3,8]$, $A_0\in[1/5, 1]$ MV/cm, and $\omega\in[5,15]$ THz. Due to the extreme time and memory cost of the numerical simulation for large systems, we are only able to generate ground-truth momentum for $N$ up to 28. In sharp contrast to the cost of numerical simulation, the training can be completed within a few GPU hours.

\begin{figure}
    \centering
    \includegraphics[width=0.8\linewidth]{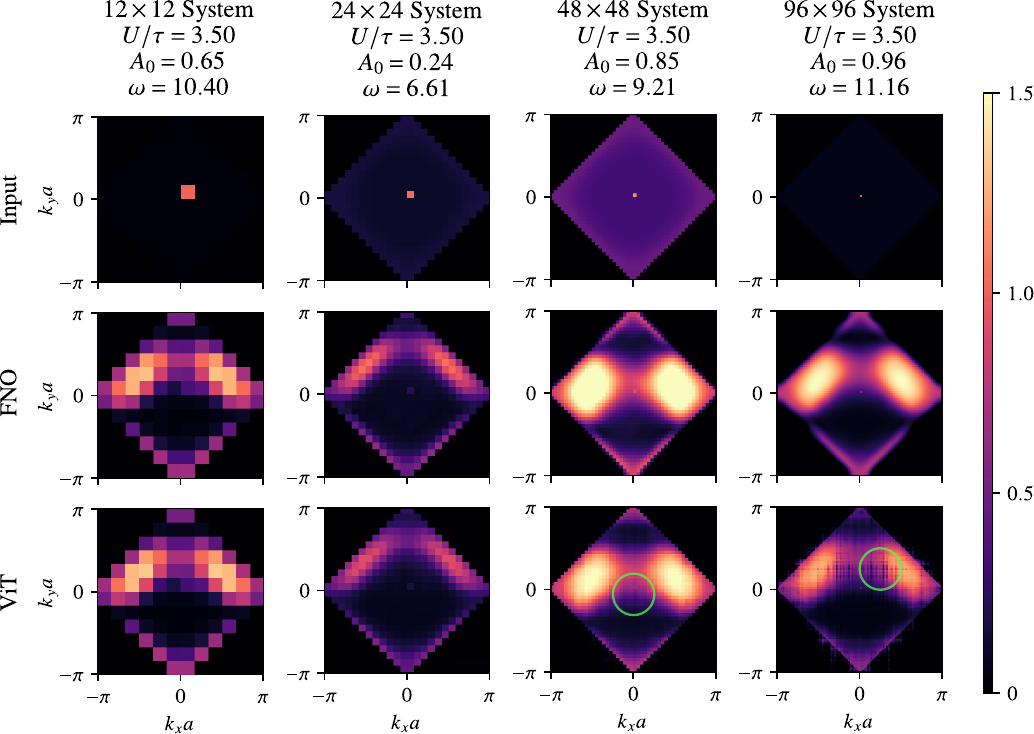}
    \caption{\@
        Example predictions of the post-pulse momentum distribution for various system sizes. All but the leftmost column are predictions on systems larger than those in the training data. All but the rightmost column are initial states that are generated using our numeric simulation. The last column is an approximation of an initial state in which the Brillouin zone is filled with a small constant value everywhere except for a target momentum excitation. The first row is the initial state at $t=0$ fs, the second row is the FNO prediction at $t=200$ fs, and the third row is the ViT prediction at $t=200$ fs. No ground truth is included, as the computation cost is intractable to run real-time evolution for the two larger systems. The ViT exhibits unphysical artifacts, circled in green.
    }\label{fig:large_inputs}
    \vspace{-1em}
\end{figure}


\paragraph{Efficient and Accurate In-distribution Prediction}
Both models are very accurate, with similar test losses under 2\% per noise realization across all system sizes that were trained on. Both models achieve an approximately 1500$\times$ speedup over the numerical simulation on system size $N=8$.

\paragraph{Toward the Thermodynamic Limit}
The focus of this work is to assess the promise of data-driven methods to target unprecedented system sizes. The inputs and predictions of both models are in \Cref{fig:large_inputs}. More examples can be found in \Cref{apdx_more_visul}. FNO can produce physically reasonable outputs. In contrast, there are noticeable unphysical artifacts in ViT predictions, e.g., the strips and dark spots circled in \Cref{fig:large_inputs}. Moreover, the complexity of FNO is $O(N^2\log N)$, which scales better than ViT with complexity $O(N^4)$ (the patch size of ViT is $O(1)$ due to the reason mentioned above).
The largest system size that our ViT can evaluate is $\sim 150$, whereas the FNO model can evaluate on a $2048\times 2048$ system within a GPU minute.

Since we are targeting $N$ where no ground truth is available, we use the average momentum density, which has a theoretical description of the behavior as $N\to\infty$, as an indirect validation (see \Cref{fig:figure_1}). The average momentum density should steadily approach a limiting value as $N\to\infty$, which we observe clearly for the FNO. For ViT, however, as shown in \Cref{fig:figure_1} (b), the observable drastically diverges from the original trend of convergence after $N=20$. Recall that 20 is the largest system size in the training dataset, which indicates that the ViT has merely memorized the training dataset, thus achieving accurate in-distribution predictions but failing to extract the underlying physics, which leads to unphysical outputs in an extrapolation setting with $N>20$ and $N\to\infty$.
We also examine the convergence in \Cref{fig:figure_1}(c).
In our setting with noisy initialization and a non-equilibrium state, the curve of FNO exhibits a steady power-law scaling after excluding the small systems with strong finite-size effects.

\textbf{Ablation Study} We conduct ablation studies for different patch sizes of ViT and different ways to encode information of system size.





\paragraph{Discussion}
Our results indicate that even when trained only with data from small system sizes, FNO has the potential to extrapolate to unprecedentedly large system sizes, targeting the thermodynamic limit in strongly correlated lattice systems. A better understanding of the underlying mechanism needs more effort. 
In this work, training is purely supervised and data-driven. Incorporating observable-based regularization at larger system sizes would be a natural and potentially effective extension.
Besides comparing to observables with known scaling, benchmarking model predictions with physical experimental observations is an important future step to validate this method.

\begin{ack}

\end{ack}

\bibliographystyle{unsrt}
\bibliography{main}

\clearpage
\appendix
\section*{Appendix}
The structure of the appendix is as follows.

\begin{itemize}
    \item \Cref{apdxA:rel_work} lists related works.
    \item \Cref{apdx:phy} provides more physical background and motivations for this research.
    \item \Cref{apdx_imple} describes the implementation details.
    \item \Cref{apdx_more_visul} provides comprehensive experimental results, some of which are not able to appear in the main text due to the page limit.
    \item \Cref{apdx_abl} presents results for ablation studies.
\end{itemize}



\section{Related Work}\label{apdxA:rel_work}

Machine learning models have shown promising results in quantum-related problems~\cite{pfau2020ab,li2023forward,liu2024unifying,li2024dof,scherbela2025accurate,shao2024antiferromagnetic}. Neural-Network Quantum States have successfully represented many-body wavefunctions~\cite{carleo_solving_2017}, which have been applied previously to the 2D Hubbard model~\cite{gu_solving_2025}. More recently, FNOs have been used for solving the Schr\"odinger equation to simulate quantum scattering~\cite{mizera_scattering_2023}, and for evolving random quantum spin systems to times far beyond the training window~\cite{shah_fourier_2024}.

FNOs exhibit structural advantages of interest for our problem. In particular, they are capable of zero-shot super-resolution, in which the model can naturally transfer its learned mappings across system sizes that were not present in its training data. The idea of transferring lattice dynamics to larger systems has previously been performed on the Ising model with convolutional neural networks (CNNs) with successful results~\cite{efthymiou_super-resolving_2019}. However, CNNs struggle to capture long-range global interactions, such as those encountered in our work, which is a motivation for using an FNO or ViT instead.
In the area of variational Monte Carlo,~\cite{chen2024empowering} has explored the use of an RNN to transfer across different lattice sizes for ground state calculation. However, finetuning is needed and necessary in a new system size.
Recent work has shown that neural operators are capable of generalizing lattice field sampling to larger systems than those trained on~\cite{mate_multi-lattice_2024}.

\section{Physical Background and Motivation}\label{apdx:phy}

In this work, we study the nonequilibrium dynamics of strongly correlated electrons on a two-dimensional square lattice, described by the single-band Hubbard model~\cite{hubbard1963electron}. The many-body quantum state is characterized by electron occupations at each lattice site, where sites can be empty, singly occupied (spin up or down), or doubly occupied. The time-dependent Hamiltonian under external driving is:
\begin{equation}
    H(t) = -\tau\sum_{\mathbf{j},\boldsymbol{\delta},\alpha} e^{ie\boldsymbol{\delta\cdot A}(t)} c_{\mathbf{j}+\boldsymbol{\delta},\alpha}^{\dagger} c_{\mathbf{j},\alpha} + U \sum_{\mathbf{j}} n_{\mathbf{j}\uparrow} n_{\mathbf{j}\downarrow},
\end{equation}
where $c_{\mathbf{j},\alpha}^{(\dagger)}$ are fermionic annihilation (creation) operators at site $\mathbf{j}\in {\{1,2,\ldots,N\}}^2$ with spin $\alpha \in \{\uparrow,\downarrow\}$, $\boldsymbol{\delta} \in \{\pm\hat{x}, \pm\hat{y}\}$ connects nearest neighbors, $\tau$ is the hopping amplitude, and $U$ represents the on-site Coulomb repulsion. The $e$ in the exponent denotes electron charge. External optical pulses are incorporated via Peierls substitution through the time-dependent vector potential $\mathbf{A}(t)$, related to the electric field via $\vec{E}(t) = -\partial \mathbf{A}(t)/\partial t$. We employ Gaussian pulses of the form
\[A(t)=A_0\exp\left(-4\ln(2)\frac{{(t-t_0)}^{2}}{\tau_{\text{FWHM}}^2}\right)\cos(\omega(t-t_0)),\]
where $A_0$, $\omega$, and $\tau_{\text{FWHM}}$ control the amplitude, frequency, and width of the drive.

While nonlinear photoexcitation processes in the $U \gg \tau$ limit have been previously studied in depth using Gaussian states~\cite{andrei_subgap_2025}, the $U \gtrsim \tau$ case enables new carrier excitation mechanisms, and at the same time requires a more sophisticated theoretical description. Here, we focus on impact ionization and avalanche photoexcitation, whose carrier multiplication effect has been proposed~\cite{manousakis_photovoltaic} as a pathway towards designing solar cells that would outperform the conventional, semiconductor-based ones in widespread use today.

Our setting falls under the general field of ultrafast dynamics, in which driven systems traverse markedly non-equilibrium states upon optical excitation. The main difficulty in making theoretical predictions lies precisely in these pre-equilibration times, such as during and shortly after the application of an external pulse. Therefore, we focus our studies on predicting the quasi-particle population at $t = 200$ fs, which is the scale of our pulse duration. Also note that we focus on driving with low-frequency pulses, which is considerably more challenging than the high-frequency limit, where standard techniques such as Floquet theory apply. In contrast, low-frequency drives lead to resonances and strong non-adiabatic effects that are more challenging to capture.

While in principle the evolution of a quantum system can be described by the Schrödinger equation \[i\pdv{t}\ket{\psi(t)} = H(t) \ket{\psi(t)},\] in practice the exponential scaling of Hilbert space dimension with system size makes this approach intractable for all but the smallest simulations. Variational states~\cite{variational_geometry} are a powerful method for approximating the quantum evolution in a computationally tractable manner, while capturing the relevant physical processes. Since quantum fluctuations are essential for describing impact ionization, we use an ensemble of Gaussian states rather than a single variational state, sampled in the spirit of a truncated Wigner approximation. Specifically, each trajectory in the ensemble is characterized by the correlation matrix, capturing ${\left(N^2\right)}^2$ correlation functions. Fluctuation effects are incorporated via the perturbation of the ground-state correlation matrix by random noise, before beginning the real-time evolution. Starting from the Schrödinger equation, an effective equation for the dynamics of the correlation matrix can then be derived. Since the evolution involves multiplying correlation matrices, the computational complexity of this direct correlation-matrix propagation scales as $O\left(N^6T\right)$, where $T$ is the number of time steps.

In this work, we focus on predicting the post-pulse momentum distribution defined as
\begin{equation}
    n_\mathbf{k} (t)=\sum_{\alpha}\langle d^{\dagger}_{\mathbf{k},\alpha}(t)d_{\mathbf{k},\alpha}(t)\rangle,
\end{equation}
where $d_\mathbf{k}$ operators describe elementary excitations of a Mott insulator at the Gaussian level of approximation, and are defined in detail in~\cite{andrei_subgap_2025}; they consist of a linear superposition of $c_\mathbf{k}^\dag$, which are defined via Fourier transform: \[c_{\mathbf{k},\alpha}=N^{-1/2}\sum_{\mathbf{j}}e^{-i\mathbf{k}\cdot\mathbf{j}}c_{\mathbf{j},\alpha}\] with $\mathbf{k}$ spanning the first Brillouin zone. Such momentum distributions are measurable in time- and angle-resolved photoemission spectroscopy (ARPES), capturing both equilibrium features (e.g., Fermi surface, quasiparticle renormalization) and nonequilibrium dynamics (e.g., population transfer, Floquet bands)~\cite{boschini2024time,kutnyakhov2020time}. They also serve as the central input for many downstream tasks, including transport, lifetime estimates, and phase diagnostics~\cite{valli2012correlation,lee2006doping,imada1998metal,datta1997electronic}.

\section{Implementation Details}\label{apdx_imple}

As input, the networks accept the initial momentum distribution at time $t=0$, as well as four parameters: the $U/\tau$ ratio, the reciprocal system size $1/N$ (where the system is $N\times N$), the pulse amplitude $A_0$, and the pulse frequency $\omega$. This is achieved by first broadcasting each parameter across the momentum lattice, producing homogeneous $N\times N$ arrays. These parameter arrays are then concatenated with the initial distribution and the positional embeddings to form a 7-channel tensor of shape $(7, N, N)$. The specific positional embeddings used differ between the models and are thus explained in the following sections.

\paragraph{Fourier Neural Operator}
Unlike traditional neural networks, neural operators learn mappings between function spaces. Fourier Neural Operators have proven effective for solving PDEs and can efficiently learn global interactions.
We learn the time evolution of the momentum distribution via a Fourier Neural Operator:
\begin{equation}
    \mathcal{G}_{\mathrm{FNO}}: \left(n\left(\textbf{k},0\right),\eta,\texttt{PE}_{\mathrm{FNO}}\left(\textbf{k}\right)\right) \mapsto \left(n\left(\textbf{k},T\right),\eta,\texttt{PE}_{\mathrm{FNO}}\left(\textbf{k}\right)\right),
\end{equation}
where $\eta = \left(U/\tau,A_0,\omega,1/N\right)$ are the system and pulse parameters, $\textbf{k}=\left(k_{x},k_{y}\right)\in{\left(-\pi,\pi\right]}^2$ are the allowed momenta in the Brillouin zone,  $\texttt{PE}_{\mathrm{FNO}}\left(\textbf{k}\right)$ denotes the grid positional embedding of the momentum grid, and $T=200$ fs.

The grid positional embedding is defined as:
\begin{equation}
    \texttt{PE}_{\mathrm{FNO}}\left(\textbf{k}\right)=\left(\frac{k_{x}+\pi}{2\pi},\frac{k_{y}+\pi}{2\pi}\right)\in{(0,1]}^2
\end{equation}
Our FNO is trained with 6 Fourier layers, $(10,10)$ modes, 114 hidden channels, 7 input and output channels, and lifting and projection channel ratios of 4. We use a grid embedding over ${[0, 1]}^2$ on the two momentum axes, and we use an MLP layer after each FNO block. We use the AdamW optimizer with a learning rate of $5\times10^{-4}$ and a weight decay of $10^{-4}$, and we use the ReduceLROnPlateau scheduler over 300 epochs with a learning rate reduction factor of 0.5, a patience of 15, a relative threshold of $10^{-4}$, a cooldown of 1, and a minimum learning rate of $10^{-6}$.

\paragraph{Vision Transformer}
We learn the time evolution of the momentum distribution via a Vision Transformer:
\begin{equation}
    \mathcal{G}_{\mathrm{ViT}}: \left(n\!\left(\textbf{k},0\right),\eta,\texttt{PE}_{\mathrm{ViT}}\!\left(\textbf{k}\right)\right) \mapsto \left(n\!\left(\textbf{k},T\right),\eta,\texttt{PE}_{\mathrm{ViT}}\left(\textbf{k}\right)\right),
\end{equation}
where $\eta = \left(U/\tau, A_0, \omega, 1/N\right)$ are the system and pulse parameters, $\textbf{k}=\left(k_{x},k_{y}\right)\in{\left(-\pi,\pi\right]}^2$ are the allowed momenta in the Brillouin zone, $\texttt{PE}_{\mathrm{ViT}}\left(\textbf{k}\right)$ denotes the 2D sine–cosine positional embedding of the momentum grid, and $T=200$ fs. We use the following 128-dimensional sine-cosine positional embedding, defined by:
\begin{equation}
    \texttt{PE}_{\mathrm{ViT}}\left(\textbf{k}\right)=\left[\psi(p_x),\psi(p_y)\right]\in\mathbb{R}^{128},
\end{equation}
where $(p_x, p_y)=\texttt{PE}_{\mathrm{FNO}}(\textbf{k})\in{(0,1]}^2$ are the normalized grid coordinates, and
\begin{equation}
    \psi(p_i)=\left(\sin(p_i\omega_1),\cos(p_i\omega_1),\ldots,\sin(p_i\omega_{32}),\cos(p_i\omega_{32})\right)
\end{equation}
for each $p_i\in\left\{p_x,p_y\right\}$ and with $\omega_j=10^{\frac{4}{31}(j-1)}$ where $j=1,\ldots,32$.

Our ViT is trained with an 8-layer transformer encoder, 8 attention heads, GELU activations, and an MLP expansion ratio of 2. Each token consists of a momentum state and parameters $\eta = \left(U/\tau,A_0,\omega,1/N\right)$, projected from 7 channels to 128 dimensions with a linear layer, and augmented with a 2D sinusoidal positional embedding over ${[0,1]}^2$. It is then passed through the encoder and projected back to 7 channels with a linear head. We use an AdamW optimizer with a learning rate of $10^{-3}$, weight decay $10^{-4}$, gradient-norm clipping at 1, and a CosineAnnealingLR scheduler over 500 epochs.

\newpage
\section{More Experimental Results}\label{apdx_more_visul}

\begin{figure}[h]
    \centering
    \includegraphics[width=0.85\linewidth]{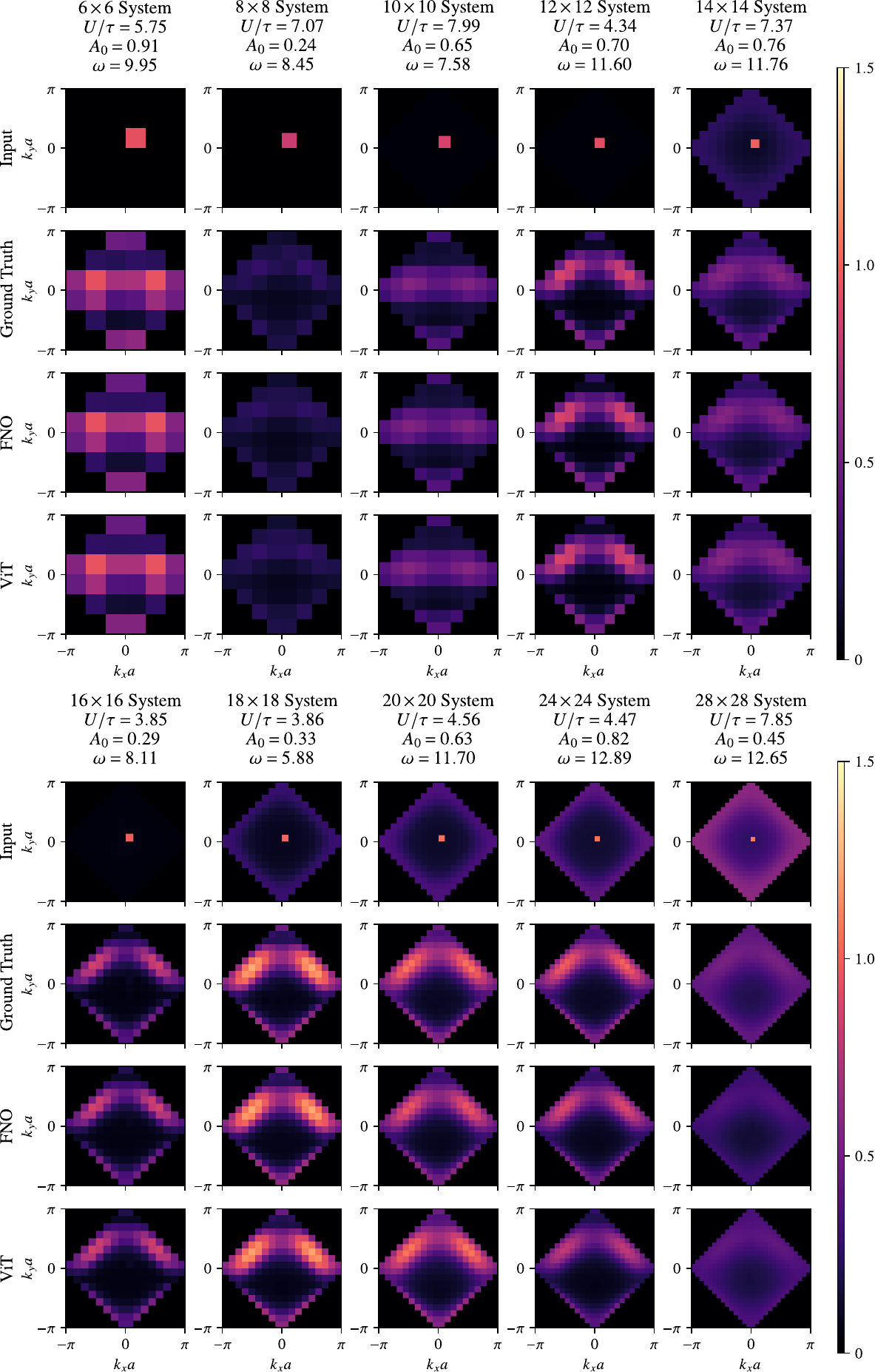}
    \caption{Randomly-selected post-pulse momentum distribution predictions on system sizes where the ground truth is accessible. Both models accurately predict the ground truth within the range of system and pulse parameters that were trained on.}\label{fig:inference_with_gt}
\end{figure}

\section{Ablation Studies}\label{apdx_abl}

\begin{figure}
    \centering
    \centering
    \includegraphics[width=0.5\textwidth]{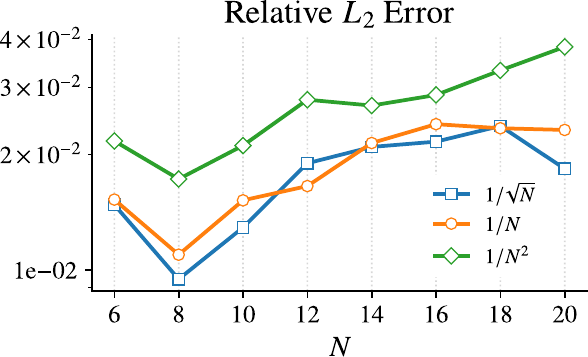}

    \caption{
        Relative $L_2$ test loss at various system sizes when the model is provided $1/\sqrt{N}$, $1/N$, or $1/N^2$ as input. Training was performed for 100 epochs with a scheduler patience of 5, and all other hyperparameters were the same.
    }\label{fig:fno_ablation_error}
\end{figure}

\begin{figure}
    \centering
    \includegraphics[width=0.98\linewidth]{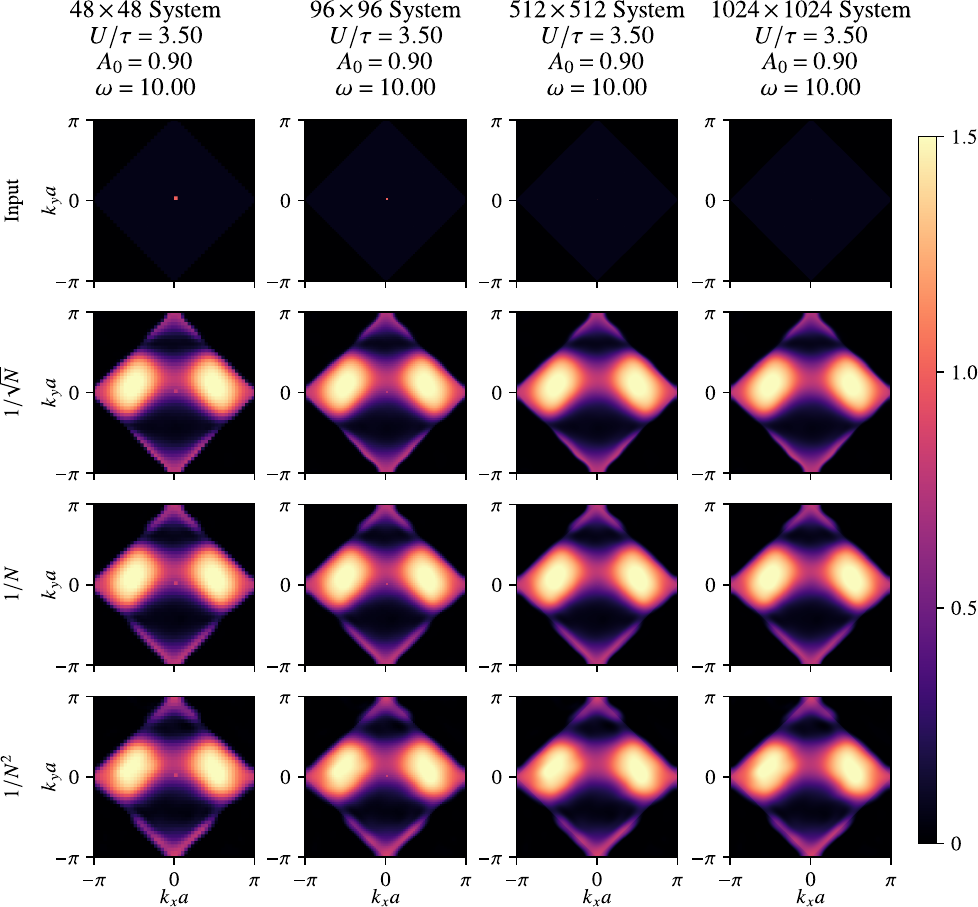}
    \caption{Example predictions on various system sizes when the model is provided $1/\sqrt{N}$, $1/N$, and $1/N^2$ as input. The predictions are very similar.}
    \label{fig:fno_ablation_distributions}
\end{figure}

\begin{figure}
    \centering
    \includegraphics[width=0.98\linewidth]{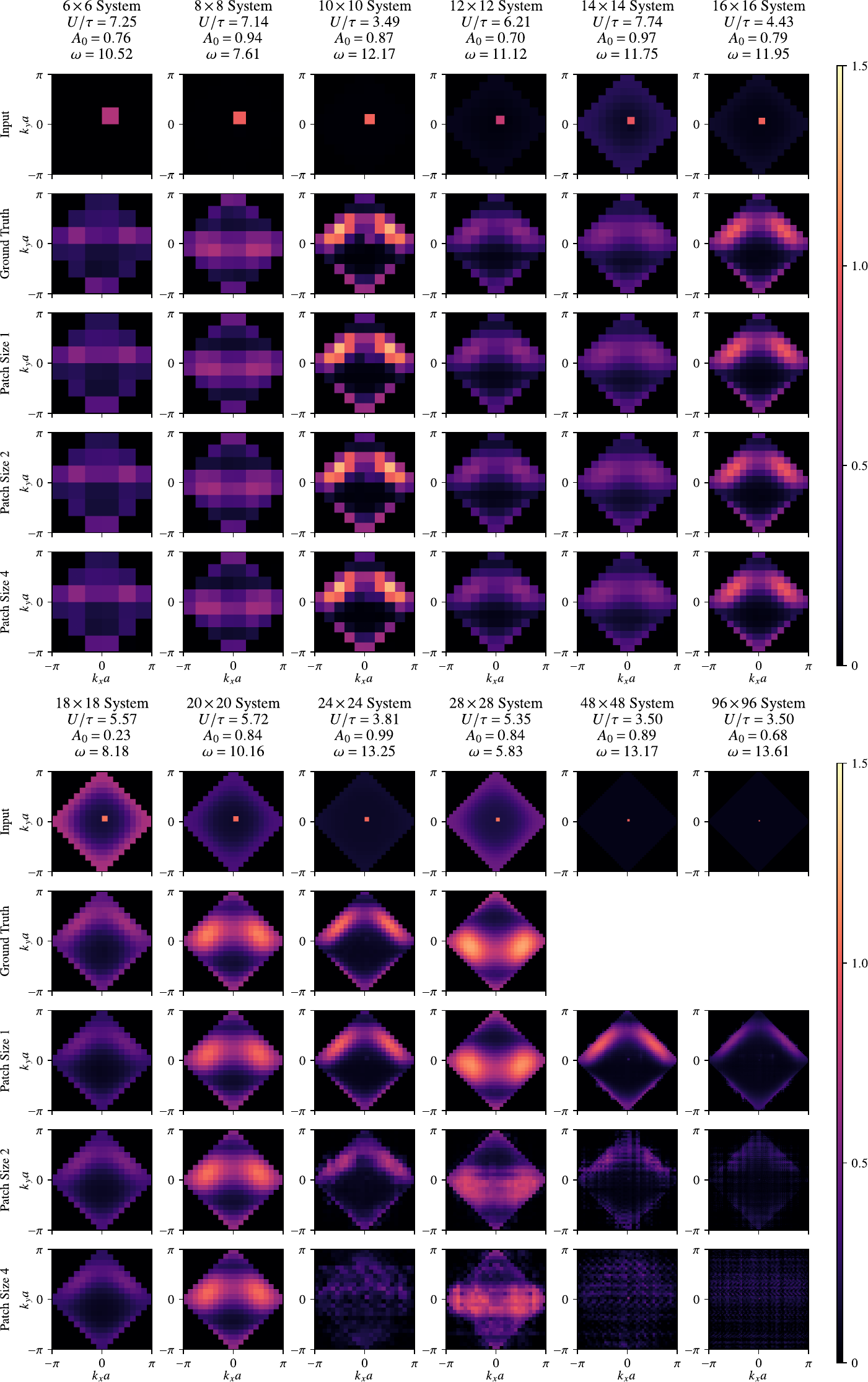}
    \caption{ViT predictions with $P\times P$-pixel patches of size $P=1, 2, 4$. The structure of the momentum distribution is maintained at large $N$ with $P=1$, and degrades with larger $P$.}\label{fig:vit_ablation}
\end{figure}

We conduct two ablation studies to understand the effect of the system size representation provided to the FNO and the patch size of the ViT on performance. In our first study, we provide the model with $1/\sqrt{N}$, $1/N$, and $1/N^2$. The final relative $L_2$ error for each model across system sizes is shown in \Cref{fig:fno_ablation_error}, and examples of model evaluations on large systems are shown in \Cref{fig:fno_ablation_distributions}. All three model predictions and losses are very similar, but the model using $1/N^2$ performs slightly worse. In our second study, we experiment with using non-overlapping $P\times P$ pixel patches for our ViT, with $P=1,2,\text{ and }4$. Because patches of size $P\times P$ for $P>2$ cannot always partition an $N\times N$ momentum distribution ($N/P$ may not be an integer), we zero-pad the momentum distribution such that its dimensions are divisible by $P$. Our predicted momentum distributions using each model, along with their associated ground truths if available, are shown in \Cref{fig:vit_ablation}. Using individual pixels as tokens $(P=1)$ proves to retain structure the best, while increasing the patch size quickly degrades the quality of the momentum distribution reconstruction at large system sizes.

\end{document}